%% file: paper.tex
\documentclass[runningheads]{llncs}
\usepackage[T1]{fontenc}
\usepackage{microtype}

\usepackage[american]{babel}
\usepackage{listings}
\usepackage{placeins}
\usepackage{graphicx}
\usepackage{color}
\usepackage{hyperref}
\usepackage{epstopdf}
\usepackage{caption}
\usepackage{subcaption}
\usepackage[table, dvipsnames, xcdraw]{xcolor}
\usepackage{longtable}
\usepackage{booktabs}
\usepackage[group-separator={,}, group-minimum-digits=4]{siunitx}
\usepackage{xspace}
\usepackage{array}
\usepackage[normalem]{ulem}
\useunder{\uline}{\ul}{}
\usepackage{orcidlink}
\usepackage{todonotes}
\usepackage{multirow}
\usepackage{float}
\usepackage{enumitem}
\usepackage{makecell}
\usepackage{svg}
\usepackage{etoolbox}
\patchcmd{\paragraph}{\itshape}{\bfseries\boldmath}{}{}

\definecolor{codegreen}{rgb}{0,0.6,0}
\definecolor{codegray}{rgb}{0.5,0.5,0.5}
\definecolor{codepurple}{rgb}{0.58,0,0.82}
\definecolor{backcolour}{rgb}{0.95,0.95,0.92}

\lstdefinestyle{mystyle}{
    commentstyle=\color{codegreen},
    keywordstyle=\color{magenta},
    numberstyle=\tiny\color{codegray},
    stringstyle=\color{codepurple},
    basicstyle=\ttfamily\footnotesize,
    breakatwhitespace=false,
    breaklines=true,
    captionpos=b,
    keepspaces=true,
    numbers=left,
    numbersep=5pt,
    showspaces=false,
    showstringspaces=false,
    showtabs=false,
    tabsize=2
}
\usepackage{diagbox}
\usepackage{amsmath}
\usepackage{amssymb}

\usepackage{mathtools, nccmath}
\usepackage{wrapfig}

\begin{document}

\title{Work Stealing for the 2D-Mesh Topology\\of Satellite Constellations in Low Earth Orbit}
%
\titlerunning{WS for the 2D-Mesh Topology in SEC}
%
\author{
    Mia Reitz\,\inst{1}\,\orcidlink{0009-0000-6188-3693}
    \and
    Dorian Chenet\,\inst{2}
    \and
    Jonas Posner\,\inst{3}\,\orcidlink{0000-0002-6491-1626}
}
\authorrunning{Reitz et al.}
%
\institute{
    University of Kassel, Kassel, Germany\\
    \email{mia.reitz@uni-kassel.de}
    \and University of Rennes, Rennes, France\\
    \email{dorian.chenet@univ-rennes.fr}
    \and Fulda University of Applied Sciences, Fulda, Germany\\
    \email{jonas.posner@cs.hs-fulda.de}
}

\hypersetup{
    pdftitle={{Work Stealing for the 2D-Mesh Topology of Satellite Constellations in Low Earth Orbit}},
    pdfsubject={AMTE26},
    pdfauthor={Mia Reitz, Dorian Chenet, and Jonas Posner},
    pdfkeywords={{Asynchronous Many-Task (AMT), ItoyoriFBC, Space Edge Computing}}
}
\maketitle              
\setcounter{footnote}{0}
\interfootnotelinepenalty=10000

\begin{abstract}
    \input{00abstract}
    \keywords{
        Asynchronous Many-Task~(AMT) \and
        Space Edge Computing~(SEC) \and
        Work Stealing \and
        ItoyoriFBC
    }
\end{abstract}
\sloppy

\input{01introduction}
\input{02background}
\input{03strategy}
\FloatBarrier
\input{04experiments}
\FloatBarrier
\input{05relatedwork}
\FloatBarrier
\input{06conclusions}
\FloatBarrier

\begin{credits}
\subsubsection*{\ackname}
This research was partially funded by the Deutsche Forschungsgemeinschaft (DFG, German Research Foundation) under project numbers~512078735 and~558599020, and the HARMONY European project (Grant Agreement ID:~101072798).

The authors gratefully acknowledge the computing time provided to them on the Goethe-NHR cluster at the Frankfurt Center for Scientific Computing.

\subsubsection*{\discintname}
The authors have no competing interests to declare that are relevant to the content of this article.
\end{credits}

\bibliography{bibo}
\end{document}

%% file: 00abstract.tex
Asynchronous Many-Task~(AMT) is a parallel programming model used in High Performance Computing~(HPC).
An AMT runtime can distribute fine-grained tasks across processing units called workers, through work stealing: when a worker has no tasks left to process, it tries to steal tasks from other workers.
Workers are not restricted to a single compute node but can also be distributed across multiple nodes of an HPC cluster.
Existing AMT runtimes assume a fully connected network with low, uniform latency and perform \textit{global} work stealing, selecting another worker at random from all workers in the system.

Space Edge Computing~(SEC) uses constellations of satellites in Low Earth Orbit~(LEO) as distributed compute clusters.
Unlike HPC clusters, LEO satellites communicate through inter-satellite links that form a sparse mesh topology.
Reaching a distant satellite requires multiple hops, each adding latency.

As a step toward adapting AMT to SEC, this paper proposes a neighbor-only work stealing strategy in which workers steal exclusively from directly connected neighbors, avoiding multi-hop communication.
An analytical model shows that restricting stealing this way yields a per-attempt latency advantage that grows with constellation size.
Preliminary experiments on an HPC cluster with an emulated mesh over uniform low-latency links isolate the effect of victim selection: the neighbor-only strategy performs within~$\pm 2.2\,\%$ of global stealing on both balanced and irregular workloads, indicating that restricting the victim set does not harm load balancing in this setting.
Taken together, the experiments suggest that neighbor-only stealing can be on a par with global stealing, and the model suggests that neighbor-only stealing becomes preferable at scale.

%% file: 01introduction.tex
\section{Introduction}\label{sec:introduction}

Reusable rockets have reduced the cost of launching satellites into Low Earth Orbit~(LEO), making large satellite constellations feasible.
Earth-observation satellites generate large amounts of data, but 
ground station links have limited throughput, so sending all raw data to Earth for processing creates a bottleneck.
Processing data on the satellites reduces the volume of data sent to the ground and delivers results faster.

Current constellations such as Starlink connect satellites through Inter-Satellite Links~(ISLs)~\cite{garcia2025direct}.
Because optical ISLs use narrow laser beams that must be precisely aimed at a single target, and each satellite carries only a few ISL terminals, ISLs connect only neighboring satellites, forming a sparse mesh topology.
Using satellites as interconnected compute nodes is called Space Edge Computing~(SEC)~\cite{SpaceEdgeEuropar24}.
Radiation-hardened commercial processors now allow satellites to run general-purpose software~\cite{mousist2025real}, turning them into programmable compute nodes.

Typical SEC workloads include Earth-observation tasks such as processing imagery:
one satellite captures an image and submits it for processing, the runtime distributes the computation across the constellation, and the result is returned to a ground station without sending the raw data to Earth first.
Other use cases include combining measurements from multiple satellites and real-time monitoring of weather or maritime traffic.
These workloads are naturally expressed as many dynamically generated subtasks, making AMT a plausible runtime model in SEC.

Asynchronous Many-Task~(AMT) is a parallel programming model widely used in High Performance Computing~(HPC)~\cite{ClaudiaTut}.
In AMT, programmers split the computation into many fine-grained sub-computations called tasks~\cite{CilkWS}.
An AMT runtime system then dynamically assigns these tasks to available processes called workers and balances the load through work stealing.
When a worker runs out of tasks, it becomes a thief and tries to take a task from another worker called a victim.
Existing AMT runtimes perform \textit{global stealing}: a thief selects a victim uniformly at random from all workers.
This assumes a fully connected network with low, uniform latency, which holds for traditional HPC clusters with high-speed interconnects such as InfiniBand.

In this work, we consider AMT to run parallel computations across SEC satellites.
In a LEO constellation, global stealing requires multi-hop communication over high-latency ISLs.
We propose a \textit{neighbor-only} stealing strategy that restricts victim selection to directly connected neighbors, ensuring single-hop communication.
In addition, we derive a condition under which neighbor-only stealing outperforms global stealing.

Our contributions are as follows:
\begin{enumerate}
    \item A neighbor-only work stealing strategy for AMT runtimes on mesh topologies.
    \item An analytical model that quantifies the per-attempt latency advantage of neighbor-only stealing as a function of constellation size.
    \item A preliminary experimental evaluation on up to 640~cores with an emulated mesh topology on a uniform low-latency HPC interconnect, which isolates the effect of victim selection from ISL latency, using the ItoyoriFBC AMT runtime~\cite{MiaFutureSideEffects}, a variant of Itoyori~\cite{itoyori}.
\end{enumerate}
This differs from prior topology-aware stealing schemes because we remove the global fallback entirely and keep all steal communication single-hop.
Our analytical model suggests that neighbor-only stealing outperforms global stealing at scale.
Moreover, our preliminary experiments show that the neighbor-only strategy performs within~$\pm 2.2\,\%$ of global stealing for both regular and irregular workloads with up to 640~cores.

The remainder of this paper is organized as follows:
Section~\ref{sec:background} provides background on SEC and AMT.
Our neighbor-only strategy is detailed in Section~\ref{sec:neighboronly}, followed by the experimental evaluation in Section~\ref{sec:experiments}.
We then discuss related work in Section~\ref{sec:relatedwork} and conclude in Section~\ref{sec:conclusions}.

%% file: 02background.tex
\section{Background}\label{sec:background}

\subsection{Space Edge Computing}\label{subsec:sec}

\begin{figure}[t]
    \centering
    \includegraphics[width=0.55\textwidth]{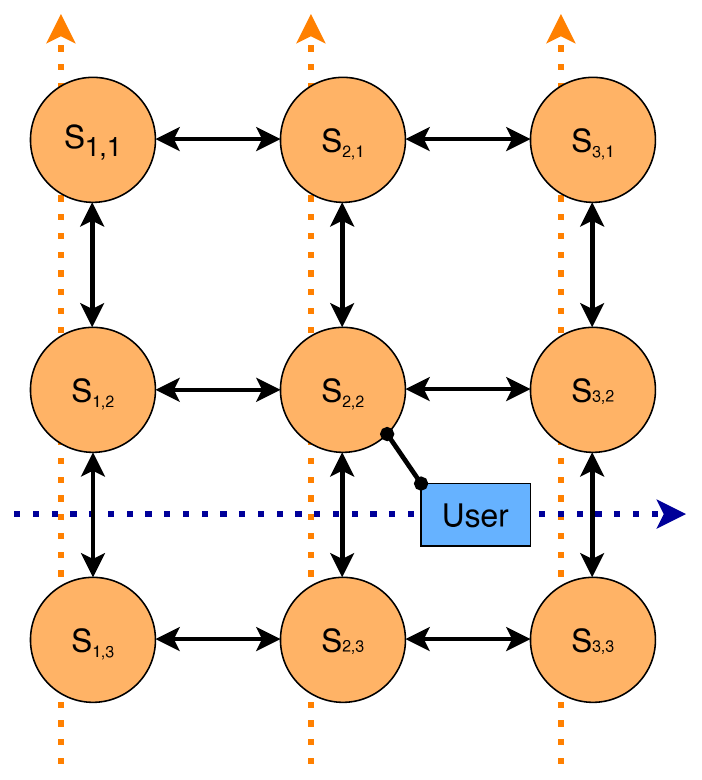}
    \caption{2D mesh topology formed by optical ISLs in a LEO constellation.
    Dashed lines show orbital paths.
    Solid lines show ISLs.
    The User represents an external entity (e.g., a ground station) that submits work to the constellation.}
    \label{fig:AMT-const}
\end{figure}

A LEO constellation consists of satellites organized in multiple orbital planes, where each plane is a ring of satellites following the same circular path around the Earth.
Each orbital plane passes through the center of the Earth, so two planes always intersect and satellites in different planes move relative to each other.
The choice of ISL technology determines the network topology.
Radio-frequency~(RF) ISLs~\cite{radhakrishnan2016survey} can reach multiple satellites at once, but data rates are limited to a few hundred Mb/s.
Optical ISLs~\cite{tiwari2020review} use laser beams and offer Gb/s data rates, but each link connects exactly two satellites.
In this work, we assume optical ISLs because they offer higher data rates and lower latency than RF links.
Each satellite maintains one link to the preceding and one to the following satellite in its orbital plane, plus one link to the nearest satellite in each of the two adjacent planes, giving four links per satellite.
This connectivity forms a 2D mesh, as shown in Figure~\ref{fig:AMT-const}.
Some constellations add wrap-around connectivity so that the satellites in each orbital plane form a ring.
SEC constellations may not need full Earth coverage, so wrap-around is not always present.
The User in Figure~\ref{fig:AMT-const} represents an external entity that submits work to the constellation, such as a ground station, a ship, or another satellite.

Running AMT programs on such a LEO constellation raises three main challenges.
First, ISL latency scales with the number of hops, making communication between distant nodes expensive~\cite{soret2020latency}.
Second, orbital planes are tilted relative to each other, so the distance between satellites in adjacent planes changes as they orbit.
This affects link latency and can temporarily break connections~\cite{yang2023interruption}.
Thus, the runtime must adapt to a changing topology.
Third, satellites can fail due to hardware faults, radiation, or power loss when entering the shadow of Earth, and limited onboard storage makes traditional recovery mechanisms impractical.

These challenges shape the design space of any runtime mechanism deployed on a constellation.
Section~\ref{sec:relatedwork} discusses how existing AMT mechanisms address them.

\begin{lstlisting}[caption={Pseudocode of a recursive Fibonacci calculation in a nested fork-join program.}, label={lst:fib}, basicstyle=\ttfamily\small, frame=single, morekeywords={spawn,sync}, float=t]
int result = fib(n);

int fib(int n) {
  if (n < 2) return n;
  int x, y;
  x = spawn fib(n - 1);
  y = fib(n - 2);
  sync;
  return x + y;
}
\end{lstlisting}

\subsection{Asynchronous Many-Task~(AMT)}\label{subsec:amt}
In this work, we consider the ItoyoriFBC AMT~\cite{MiaFutureSideEffects}, which is a variant of the Itoyori AMT runtime~\cite{itoyori}.
ItoyoriFBC is written in C++ and uses the Message Passing Interface~(MPI)~\cite{MPI} for communication between workers.

In ItoyoriFBC, each worker maintains a double-ended queue, called a deque, of tasks.
When a running task creates a child task, the worker places the remaining parent work on top of its deque and begins executing the child.
On completion, the worker pops the next task from the top of the deque.
Once the deque becomes empty, the worker becomes a thief and attempts to steal a task from the \textit{bottom} of a victim's deque.

We distinguish three phases of work stealing:
the \textit{initial phase}, where most workers are idle and tasks begin to spread;
the \textit{steady phase}, where most workers are busy and stealing balances the load;
and the \textit{termination phase}, where only a few workers still hold tasks and most steal attempts fail.

AMT programs often deploy the nested fork-join model, in which a task can create child tasks and wait for all of them to finish before continuing.
Because tasks generate new tasks during execution, the total amount of work is \textit{not} known in advance.
Listing~\ref{lst:fib} shows a recursive Fibonacci calculation as an example.
The program begins with the root task in Line~1.
The \texttt{spawn} keyword in Line~6 creates a child task \texttt{fib(n-1)} and places the remaining work of the parent task onto the deque, making it available for other workers to steal.
When the child finishes, the result is written back to~\texttt{x} in the parent task.
The \texttt{sync} keyword in Line~8 waits for all child tasks to complete.
When the root task finishes, \texttt{result} holds the final value.

%% file: 03strategy.tex
\section{Neighbor-Only Stealing}\label{sec:neighboronly}

\subsection{Algorithm}\label{subsec:algorithm}

AMT runtimes typically perform \textit{global stealing}, selecting victims uniformly at random from all workers~\cite{CilkWS}.
We propose the \textit{neighbor-only} strategy, which restricts victim selection to directly connected neighbors.

A thief executes the following algorithm during work stealing:
\begin{enumerate}
    \item Identify the current set of neighbors. In our experiments, this set is precomputed at initialization, but in a real SEC deployment it would be updated as satellites move and the topology changes.
    \item Select a neighbor~$u$ uniformly at random.
    \item Attempt to steal from~$u$.
    \item If the steal fails, repeat from step~1.
\end{enumerate}

Figure~\ref{fig:topology_concept} contrasts global and neighbor-only stealing on a $5\times5$ mesh.
The global strategy may need several hops to reach a victim, while a neighbor-only steal always completes in a single hop.

\subsection{Implementation}\label{subsec:implementation}
We implemented our neighbor-only strategy by extending ItoyoriFBC.
The source code is publicly available on Zenodo~\cite{ReitzCode2026}.

To emulate a LEO constellation on a fully connected HPC cluster, we restricted victim selection to a 2D grid overlay:
At initialization, each process is assigned a coordinate in a 2D grid.
Our extended ItoyoriFBC runtime pre-calculates the list of neighbors for each process based on this grid.
During work stealing, the runtime selects victims exclusively from this neighbor list.

In nested fork-join programs, a child task must return its result to the parent task (see Listing~\ref{lst:fib}, Line~6), which requires a single message on a fully connected network but may traverse multiple hops on a mesh.

We did not modify result propagation in ItoyoriFBC, so results are sent directly between workers, bypassing the grid topology.
Any performance difference in our experiments is therefore due to victim selection alone.
We expect result messages under neighbor-only stealing to travel short distances on average, since tasks are stolen from directly connected neighbors.

\input{03strategy_tikz}

\subsection{Analysis}\label{subsec:analysis}

We now compare the two strategies analytically, first for the steady phase and then for the initial and termination phases.
The analysis rests on the following assumptions:
\begin{itemize}
    \item[(i)] The constellation forms a $\sqrt{N}\times\sqrt{N}$ 2D mesh with $N$~nodes, each having four neighbors (the boundary is discussed below).
    \item[(ii)] Each hop costs a fixed single-hop ISL latency~$\tau$, messages follow the shortest paths, and there is no congestion.
    \item[(iii)] Steal attempts are independent and each attempt takes the round-trip time between thief and victim regardless of outcome.
\end{itemize}

\paragraph{Steady Phase.}
In the steady phase, most workers are busy and work is spread across the mesh.
For neighbor-only stealing, the round-trip time is constant at $2\tau$.
For global stealing, the average number of hops between two random nodes in a $\sqrt{N}\times\sqrt{N}$ mesh is $\tfrac{2}{3}\sqrt{N}$, giving a round-trip time of $\tfrac{4}{3}\sqrt{N}\tau$.

Let $P_{neighbor}$ and $P_{global}$ denote the probability that a randomly chosen victim has stealable work under the neighbor-only and global strategies, respectively.
The expected time to acquire a task, $\mathbb{E}[T]$, is the product of the cost per attempt and the expected number of attempts.
A steal attempt is successful with probability~$P$, so the expected number of attempts is~$1/P$:

\begin{equation}
    \mathbb{E}[T_{neighbor}] = \frac{2\tau}{P_{neighbor}} \quad , \quad \mathbb{E}[T_{global}] = \frac{\tfrac{4}{3}\sqrt{N}\tau}{P_{global}}
\end{equation}

The neighbor-only strategy is faster when $\mathbb{E}[T_{neighbor}] < \mathbb{E}[T_{global}]$.
Substituting and simplifying gives:

\begin{equation}
    \label{eq:inequality}
    \frac{P_{global}}{P_{neighbor}} < \frac{2}{3}\sqrt{N}
\end{equation}

Inequality~\eqref{eq:inequality} shows that neighbor-only stealing is faster unless global stealing finds work at least~$\tfrac{2}{3}\sqrt{N}$ times more often.
During the steady phase, most workers have tasks, so a neighbor is approximately as likely to have stealable work as any other worker.
This keeps $P_{global}/P_{neighbor}$ close to~1.
Since $\tfrac{2}{3}\sqrt{N}$ grows with constellation size, the condition is easier to satisfy for larger constellations.
For example, in a 100-node constellation, $\tfrac{2}{3}\sqrt{100} \approx 6.7$, so neighbor-only stealing is faster unless global stealing finds work more than $6.7$ times as often per attempt.

To illustrate the latency difference, Table~\ref{tab:analytical} lists the expected round-trip time for a single steal attempt under both strategies for several constellation sizes, assuming $\tau = 5\,$ms.
The neighbor-only round trip is always $2\tau = 10\,$ms, while the global round trip grows with the mesh size.

\begin{table}[ht]
    \centering
    \caption{Expected round-trip time of a single steal attempt, and the threshold $\tfrac{2}{3}\sqrt{N}$, for several constellation sizes with single-hop ISL latency $\tau = 5\,$ms.
    The threshold is the factor by which global stealing would need to find work more often per attempt to offset its higher latency.}
    \label{tab:analytical}
    \begin{tabular}{r r r r}
        \toprule
        \textbf{Nodes ($N$)} & \textbf{Threshold $\tfrac{2}{3}\sqrt{N}$} & \textbf{Neighbor RT} & \textbf{Global RT} \\
        \midrule
        25   & 3.3  & 10\,ms  & 33\,ms \\
        100  & 6.7  & 10\,ms  & 67\,ms \\
        400  & 13.3 & 10\,ms  & 133\,ms \\
        1\,600 & 26.7 & 10\,ms  & 267\,ms \\
        \bottomrule
    \end{tabular}
\end{table}

The analysis above assumes that every node has four neighbors, but boundary nodes (corners and edges) have fewer, which reduces the number of available victims.
The boundary fraction shrinks as the grid grows, so this effect becomes negligible for large constellations.

\paragraph{Initial Phase.}
At the beginning of the initial phase, all work starts on a single worker.
With global stealing, any thief in the system can target the root worker, so tasks spread to all $N$~workers within a few steal rounds.
With neighbor-only stealing, tasks spread outward from the root one hop at a time.

On a $\sqrt{N}\times\sqrt{N}$ grid, the farthest corner is at most $2(\sqrt{N}{-}1)$ hops from the root, so all workers can receive tasks within roughly $2\sqrt{N}$ steal rounds, each taking one neighbor round trip of~$2\tau$.
The initial-phase duration with neighbor-only stealing is therefore approximately $4\sqrt{N}\tau$.
For example, in a 400-satellite constellation with $\tau = 5\,$ms, this amounts to about $400\,$ms.
Since typical AMT computations are much longer, the initial phase is a small fraction of the total execution time under either strategy.

\paragraph{Termination Phase.}
During the termination phase, only a few workers still hold tasks.
A thief restricted to neighbors is less likely to find one of these remaining workers than a thief that can target any worker at random.
For balanced workloads, the few remaining tasks tend to be scattered across the mesh rather than clustered, so a neighbor-only thief is unlikely to be adjacent to a worker that still has work.
For irregular workloads, work remains distributed unevenly, so neighbors of busy workers are more likely to have stealable tasks, making neighbor-only stealing more competitive.

\paragraph{Summary.}
The steady-phase analysis predicts a per-attempt latency advantage for neighbor-only stealing that grows with $\sqrt{N}$ under realistic ISL latencies.
The initial phase adds a small overhead of order $4\sqrt{N}\tau$, and the termination phase is the only phase where the restricted victim set could penalize neighbor-only stealing, and even then only for balanced workloads.
The experiments in Section~\ref{sec:experiments} isolate the effect of victim selection from ISL latency by running on a uniform low-latency interconnect, allowing us to test the assumption that $P_{global}/P_{neighbor}\approx 1$ in practice.

%% file: 03strategy_tikz.tex
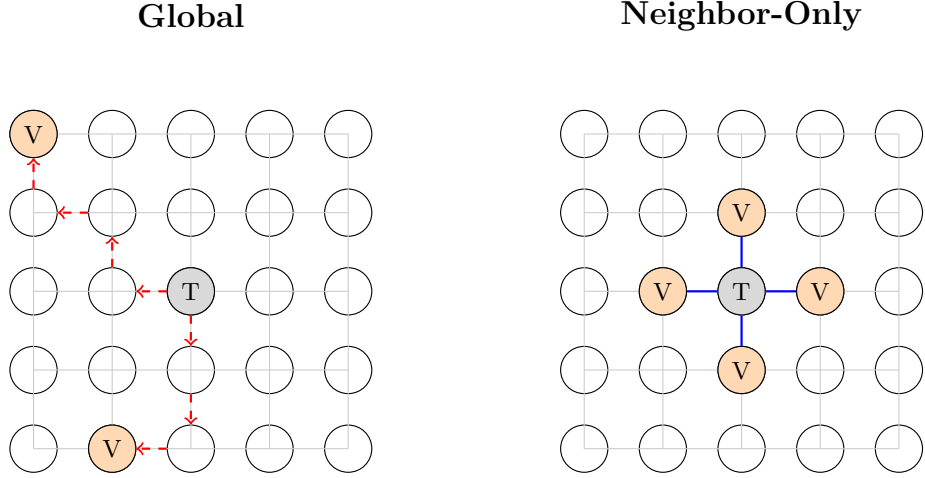
\begin{figure}[t]
    \centering
    \resizebox{\textwidth}{!}{%
    \begin{tikzpicture}[transform shape]
        \tikzstyle{sat}=[circle, draw=black, fill=white, minimum size=0.6cm, inner sep=0pt]
        \tikzstyle{thief}=[circle, draw=black, fill=gray!30, minimum size=0.6cm, inner sep=0pt]
        \tikzstyle{victim}=[circle, draw=black, fill=orange!30, minimum size=0.6cm, inner sep=0pt]
        \tikzstyle{link}=[-, draw=gray!40, thin]
        \tikzstyle{global_steal}=[->, draw=red, thick, dashed]
        \tikzstyle{local_steal}=[-, draw=blue, thick]

        \begin{scope}[xshift=0cm]
            \node at (2.0, 5.5) {\large \textbf{Global}};
            
            \foreach \x in {0,...,4}
                \foreach \y in {0,...,4}
                    \node[sat] (n\x\y) at (\x, \y) {};
            
            \foreach \x in {0,...,4}
                \foreach \y in {0,...,3}
                    \draw[link] (\x,\y) -- (\x,\y+1);
            \foreach \x in {0,...,3}
                \foreach \y in {0,...,4}
                    \draw[link] (\x,\y) -- (\x+1,\y);

            \node[thief] (t) at (2,2) {T};
            \node[victim] (v0) at (0,4) {V};
            \node[victim] (v1) at (1,0) {V};
            
            \draw[global_steal] (t) -- (n12);
            \draw[global_steal] (n12) -- (n13);
            \draw[global_steal] (n13) -- (n03);
            \draw[global_steal] (n03) -- (n04);

            \draw[global_steal] (t) -- (n21);
            \draw[global_steal] (n21) -- (n20);
            \draw[global_steal] (n20) -- (n10);
        \end{scope}

        \begin{scope}[xshift=7cm]
            \node at (2.0, 5.5) {\large \textbf{Neighbor-Only}};
            
            \foreach \x in {0,...,4}
                \foreach \y in {0,...,4}
                    \node[sat] (n\x\y) at (\x, \y) {};
            
            \foreach \x in {0,...,4}
                \foreach \y in {0,...,3}
                    \draw[link] (\x,\y) -- (\x,\y+1);
            \foreach \x in {0,...,3}
                \foreach \y in {0,...,4}
                    \draw[link] (\x,\y) -- (\x+1,\y);

            \node[thief] (t) at (2,2) {T};
            \node[victim] (v0) at (1,2) {V};
            \node[victim] (v1) at (3,2) {V};
            \node[victim] (v2) at (2,1) {V};
            \node[victim] (v3) at (2,3) {V};
            
            \draw[local_steal] (t) -- (n23);
            \draw[local_steal] (t) -- (n21);
            \draw[local_steal] (t) -- (n12);
            \draw[local_steal] (t) -- (n32);
        \end{scope}
    \end{tikzpicture}
    }
    \caption{Comparison of global (left) and neighbor-only (right) stealing on a $5\times5$ mesh.
    The thief~(T, gray) attempts to reach victims~(V, orange).
    Global stealing targets any worker, requiring multi-hop paths (dashed red arrows).
    Neighbor-only stealing targets only directly connected neighbors, using single hops (solid blue arrows).}
    \label{fig:topology_concept}
\end{figure}

%% file: 04experiments.tex
\section{Experiments}\label{sec:experiments}

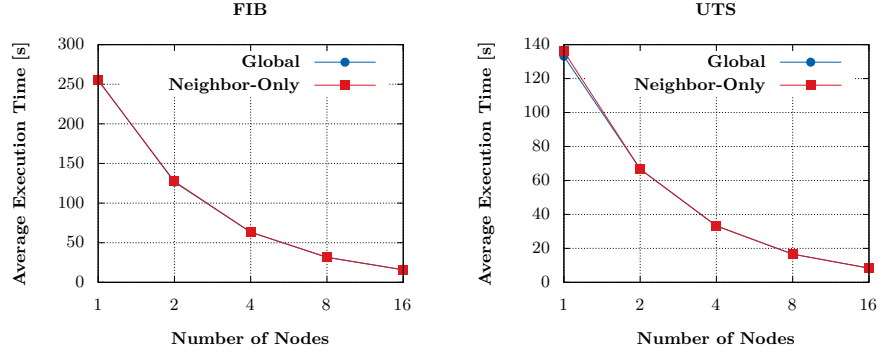
\begin{figure}[tbp]
    \centering
    \begin{minipage}{0.495\textwidth}
        \centering
        \resizebox{\linewidth}{!}{
            \input{plot_fib}
        }
    \end{minipage}
    \hfill
    \begin{minipage}{0.495\textwidth}
        \centering
        \resizebox{\linewidth}{!}{
            \input{plot_uts}
        }
    \end{minipage}

    \caption{Strong-scaling execution time (seconds, lower is better) for FIB (left) and UTS (right) on 1 -- 16 HPC nodes (40 -- 640 cores), comparing global stealing (blue) and neighbor-only stealing (red).
    Each point is the average of 10 runs.
    Curves overlap throughout, indicating that the two strategies perform roughly equivalently across all node counts.}
    \label{fig:fib_uts}
\end{figure}

\begin{table}[tbp]
    \centering
    \caption{Average execution time (seconds, 10 runs) for FIB and UTS under global and neighbor-only stealing on 1--16 HPC nodes.}
    \label{tab:absoluteTimes}
    \begin{tabular}{c rr c rr}
        \toprule
        & \multicolumn{2}{c}{\textbf{FIB}} & \phantom{a} & \multicolumn{2}{c}{\textbf{UTS}} \\
        \cmidrule{2-3} \cmidrule{5-6}
        \textbf{Nodes} & \textbf{Global} & \textbf{Neighbor-Only} && \textbf{Global} & \textbf{Neighbor-Only} \\
        \midrule
        1  & 255.44 & 254.82 && 133.18 & 136.01 \\
        2  & 126.48 & 127.30 && 66.70  & 66.64 \\
        4  & 63.18  & 63.38  && 33.24  & 33.19 \\
        8  & 31.47  & 31.68  && 16.60  & 16.61 \\
        16 & 15.77  & 15.87  && 8.37   & 8.34 \\
        \bottomrule
    \end{tabular}
\end{table}

\begin{figure}[tbp]
    \centering
    \resizebox{0.9\textwidth}{!}{
        \input{plot_diff}
    }
    \caption{Relative performance of neighbor-only versus global stealing, calculated as $(T_{neighbor} - T_{global}) / T_{global}$ in percent.
    Positive values indicate neighbor-only is slower; negative values indicate it is faster.
    Values remain within $\pm 2.2\,\%$ across all node counts and both workloads.
    }
    \label{fig:diff}
\end{figure}
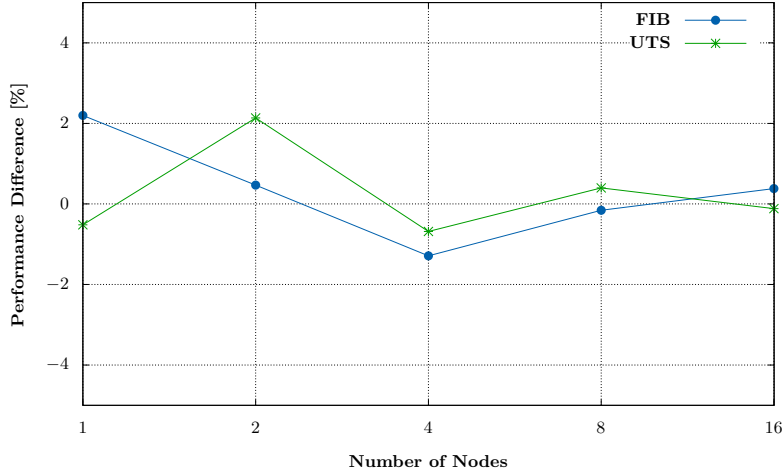

The analysis of Section~\ref{subsec:analysis} predicts that neighbor-only stealing gains a per-attempt latency advantage on a high-latency mesh, provided that the probability of finding work per attempt is comparable to global stealing.
We now test this work-finding assumption in a controlled setting: on a uniform low-latency interconnect, where any performance difference between the two strategies is attributable to victim selection alone rather than to hop latency.

\subsection{Experimental Setup}
We executed all experiments on the Goethe-NHR HPC cluster~\cite{ClusterGoethe}.
All nodes are identical and connected by a 100\,Gb/s (4X EDR) InfiniBand interconnect.
Each node is equipped with two 20-core Intel Xeon Skylake Gold 6148 CPUs and 192~GB of main memory.

We evaluated our neighbor-only strategy with two benchmarks with different task tree structures, using our extended ItoyoriFBC implementation from Section~\ref{subsec:implementation}.
\begin{itemize}
    \item \textbf{Fibonacci (FIB):} The recursive Fibonacci implementation from Section~\ref{subsec:amt}, which generates a balanced task tree.
    We set $n = 62$, and subtrees with $n \leq 32$ are computed sequentially to avoid task creation overhead.

    \item \textbf{Unbalanced Tree Search (UTS):} A benchmark that generates a highly irregular task tree by using a hash function to pseudo-randomly decide the number of children at each tree node, creating severe load imbalance~\cite{UTSPaper}.
    We configured the geometric distribution variant with an expected branching factor~$b=4$, a maximum depth~$d=16$, and an initial seed~$r=19$.
\end{itemize}
We compiled with Open~MPI~5.0.5 and g++~11.4.1 using \texttt{-O3} optimization.

We ran strong-scaling experiments on 1 to 16 nodes (40 to 640~cores).
Processes are mapped to a square 2D grid with side length $\lceil\sqrt{C}\rceil$, where $C$ is the total number of cores.
Rows are filled in order, so the last row may contain fewer processes.
Processes at the end of the last row have two neighbors, the same as any other corner process.
Each reported value is the average of 10 runs.
We do not report confidence intervals, as per-run variability was small relative to the differences.
All nodes communicate with the uniform low latency of the HPC interconnect, isolating the effect of victim selection from ISL latency.

\subsection{Results and Discussion}
Figure~\ref{fig:fib_uts} presents the strong-scaling results for both benchmarks.
The execution times for global stealing (blue) and neighbor-only stealing (red) are nearly identical across all node counts.
Both strategies scale near-linearly from~1 to 16~nodes, indicating that restricting victim selection to neighbors does not degrade scalability.
Table~\ref{tab:absoluteTimes} lists the corresponding execution times.

Figure~\ref{fig:diff} shows the relative performance, calculated as $(T_{neighbor} - T_{global}) / T_{global}$ in percent.
A positive value indicates neighbor-only stealing is slower, a negative value indicates it is faster.
The performance difference stays within $\pm 2.2\,\%$ with no consistent trend, as expected on a uniform low-latency interconnect, suggesting that the two strategies perform equivalently in these preliminary experiments.

For the balanced FIB benchmark, neighbor-only stealing is slightly slower (up to $+0.65\,\%$ at 16 nodes).
A likely cause is the termination phase: in a balanced tree, branches finish at roughly the same time, so the last tasks remain on a few scattered workers, and a thief restricted to neighbors is less likely to be adjacent to one of them.
For the irregular UTS benchmark, neighbor-only stealing occasionally outperforms global stealing (e.g., $-0.36\,\%$ at 16 nodes).
We attribute this to localized clusters of work in the irregular tree, where neighbors of a busy worker are more likely to hold stealable tasks than a uniformly random worker.

With ISL latency, each neighbor-only steal attempt would complete faster than a global one, and the analytical model from Section~\ref{subsec:analysis} predicts that this difference grows with constellation size.
To estimate the effect, consider a 400-satellite constellation with $\tau = 5\,$ms.
Table~\ref{tab:analytical} gives a global round-trip time of $133\,$ms compared to $10\,$ms for neighbor-only stealing.
Since both strategies found work at comparable rates in our experiments, the ratio $P_{global}/P_{neighbor}$ is close to~1, well below the threshold of~$13.3$.
In this scenario, each neighbor-only steal attempt would complete roughly $13{\times}$ faster than a global one.
This estimate does not account for the congestion that multi-hop steals would cause on intermediate ISLs, which would further penalize the global strategy.

Taken together, the measurements suggest no measurable load-balancing penalty from restricting victim selection in these preliminary experiments, while the analytical model projects a substantial per-attempt latency advantage at realistic ISL costs.

%% file: plot_fib.tex
\begingroup
\bfseries
  \makeatletter
  \providecommand\color[2][]{%
    \GenericError{(gnuplot) \space\space\space\@spaces}{%
      Package color not loaded in conjunction with
      terminal option `colourtext'%
    }{See the gnuplot documentation for explanation.%
    }{Either use 'blacktext' in gnuplot or load the package
      color.sty in LaTeX.}%
    \renewcommand\color[2][]{}%
  }%
  \providecommand\includegraphics[2][]{%
    \GenericError{(gnuplot) \space\space\space\@spaces}{%
      Package graphicx or graphics not loaded%
    }{See the gnuplot documentation for explanation.%
    }{The gnuplot epslatex terminal needs graphicx.sty or graphics.sty.}%
    \renewcommand\includegraphics[2][]{}%
  }%
  \providecommand\rotatebox[2]{#2}%
  \@ifundefined{ifGPcolor}{%
    \newif\ifGPcolor
    \GPcolortrue
  }{}%
  \@ifundefined{ifGPblacktext}{%
    \newif\ifGPblacktext
    \GPblacktextfalse
  }{}%
  \let\gplgaddtomacro\g@addto@macro
  \gdef\gplbacktext{}%
  \gdef\gplfronttext{}%
  \makeatother
  \ifGPblacktext
    \def\colorrgb#1{}%
    \def\colorgray#1{}%
  \else
    \ifGPcolor
      \def\colorrgb#1{\color[rgb]{#1}}%
      \def\colorgray#1{\color[gray]{#1}}%
      \expandafter\def\csname LTw\endcsname{\color{white}}%
      \expandafter\def\csname LTb\endcsname{\color{black}}%
      \expandafter\def\csname LTa\endcsname{\color{black}}%
      \expandafter\def\csname LT0\endcsname{\color[rgb]{1,0,0}}%
      \expandafter\def\csname LT1\endcsname{\color[rgb]{0,1,0}}%
      \expandafter\def\csname LT2\endcsname{\color[rgb]{0,0,1}}%
      \expandafter\def\csname LT3\endcsname{\color[rgb]{1,0,1}}%
      \expandafter\def\csname LT4\endcsname{\color[rgb]{0,1,1}}%
      \expandafter\def\csname LT5\endcsname{\color[rgb]{1,1,0}}%
      \expandafter\def\csname LT6\endcsname{\color[rgb]{0,0,0}}%
      \expandafter\def\csname LT7\endcsname{\color[rgb]{1,0.3,0}}%
      \expandafter\def\csname LT8\endcsname{\color[rgb]{0.5,0.5,0.5}}%
    \else
      \def\colorrgb#1{\color{black}}%
      \def\colorgray#1{\color[gray]{#1}}%
      \expandafter\def\csname LTw\endcsname{\color{white}}%
      \expandafter\def\csname LTb\endcsname{\color{black}}%
      \expandafter\def\csname LTa\endcsname{\color{black}}%
      \expandafter\def\csname LT0\endcsname{\color{black}}%
      \expandafter\def\csname LT1\endcsname{\color{black}}%
      \expandafter\def\csname LT2\endcsname{\color{black}}%
      \expandafter\def\csname LT3\endcsname{\color{black}}%
      \expandafter\def\csname LT4\endcsname{\color{black}}%
      \expandafter\def\csname LT5\endcsname{\color{black}}%
      \expandafter\def\csname LT6\endcsname{\color{black}}%
      \expandafter\def\csname LT7\endcsname{\color{black}}%
      \expandafter\def\csname LT8\endcsname{\color{black}}%
    \fi
  \fi
    \setlength{\unitlength}{0.0500bp}%
    \ifx\gptboxheight\undefined%
      \newlength{\gptboxheight}%
      \newlength{\gptboxwidth}%
      \newsavebox{\gptboxtext}%
    \fi%
    \setlength{\fboxrule}{0.5pt}%
    \setlength{\fboxsep}{1pt}%
    \definecolor{tbcol}{rgb}{1,1,1}%
\begin{picture}(4648.00,3968.00)%
    \gplgaddtomacro\gplbacktext{%
      \csname LTb\endcsname
      \put(888,768){\makebox(0,0)[r]{\strut{}$0$}}%
      \csname LTb\endcsname
      \put(888,1181){\makebox(0,0)[r]{\strut{}$50$}}%
      \csname LTb\endcsname
      \put(888,1594){\makebox(0,0)[r]{\strut{}$100$}}%
      \csname LTb\endcsname
      \put(888,2008){\makebox(0,0)[r]{\strut{}$150$}}%
      \csname LTb\endcsname
      \put(888,2421){\makebox(0,0)[r]{\strut{}$200$}}%
      \csname LTb\endcsname
      \put(888,2834){\makebox(0,0)[r]{\strut{}$250$}}%
      \csname LTb\endcsname
      \put(888,3247){\makebox(0,0)[r]{\strut{}$300$}}%
      \csname LTb\endcsname
      \put(1032,528){\makebox(0,0){\strut{}$1$}}%
      \csname LTb\endcsname
      \put(1828,528){\makebox(0,0){\strut{}$2$}}%
      \csname LTb\endcsname
      \put(2623,528){\makebox(0,0){\strut{}$4$}}%
      \csname LTb\endcsname
      \put(3419,528){\makebox(0,0){\strut{}$8$}}%
      \csname LTb\endcsname
      \put(4214,528){\makebox(0,0){\strut{}$16$}}%
    }%
    \gplgaddtomacro\gplfronttext{%
      \csname LTb\endcsname
      \put(3144,3064){\makebox(0,0)[r]{\strut{}Global}}%
      \csname LTb\endcsname
      \put(3144,2824){\makebox(0,0)[r]{\strut{}Neighbor-Only}}%
      \csname LTb\endcsname
      \put(228,2007){\rotatebox{-270.00}{\makebox(0,0){\strut{}Average Execution Time [s]}}}%
      \put(2623,168){\makebox(0,0){\strut{}Number of Nodes}}%
      \put(2623,3607){\makebox(0,0){\strut{}FIB}}%
    }%
    \gplbacktext
    \put(0,0){\includegraphics[width={232.40bp},height={198.40bp}]{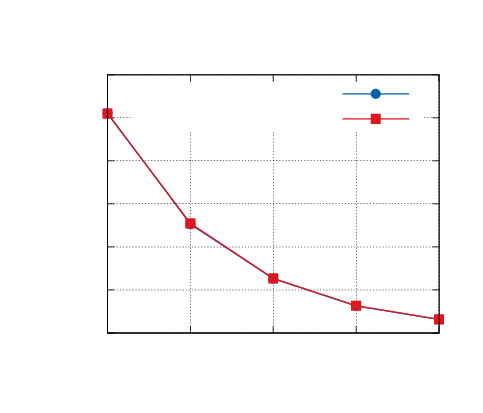}}%
    \gplfronttext
  \end{picture}%
\endgroup

%% file: plot_uts.tex
\begingroup
\bfseries
  \makeatletter
  \providecommand\color[2][]{%
    \GenericError{(gnuplot) \space\space\space\@spaces}{%
      Package color not loaded in conjunction with
      terminal option `colourtext'%
    }{See the gnuplot documentation for explanation.%
    }{Either use 'blacktext' in gnuplot or load the package
      color.sty in LaTeX.}%
    \renewcommand\color[2][]{}%
  }%
  \providecommand\includegraphics[2][]{%
    \GenericError{(gnuplot) \space\space\space\@spaces}{%
      Package graphicx or graphics not loaded%
    }{See the gnuplot documentation for explanation.%
    }{The gnuplot epslatex terminal needs graphicx.sty or graphics.sty.}%
    \renewcommand\includegraphics[2][]{}%
  }%
  \providecommand\rotatebox[2]{#2}%
  \@ifundefined{ifGPcolor}{%
    \newif\ifGPcolor
    \GPcolortrue
  }{}%
  \@ifundefined{ifGPblacktext}{%
    \newif\ifGPblacktext
    \GPblacktextfalse
  }{}%
  \let\gplgaddtomacro\g@addto@macro
  \gdef\gplbacktext{}%
  \gdef\gplfronttext{}%
  \makeatother
  \ifGPblacktext
    \def\colorrgb#1{}%
    \def\colorgray#1{}%
  \else
    \ifGPcolor
      \def\colorrgb#1{\color[rgb]{#1}}%
      \def\colorgray#1{\color[gray]{#1}}%
      \expandafter\def\csname LTw\endcsname{\color{white}}%
      \expandafter\def\csname LTb\endcsname{\color{black}}%
      \expandafter\def\csname LTa\endcsname{\color{black}}%
      \expandafter\def\csname LT0\endcsname{\color[rgb]{1,0,0}}%
      \expandafter\def\csname LT1\endcsname{\color[rgb]{0,1,0}}%
      \expandafter\def\csname LT2\endcsname{\color[rgb]{0,0,1}}%
      \expandafter\def\csname LT3\endcsname{\color[rgb]{1,0,1}}%
      \expandafter\def\csname LT4\endcsname{\color[rgb]{0,1,1}}%
      \expandafter\def\csname LT5\endcsname{\color[rgb]{1,1,0}}%
      \expandafter\def\csname LT6\endcsname{\color[rgb]{0,0,0}}%
      \expandafter\def\csname LT7\endcsname{\color[rgb]{1,0.3,0}}%
      \expandafter\def\csname LT8\endcsname{\color[rgb]{0.5,0.5,0.5}}%
    \else
      \def\colorrgb#1{\color{black}}%
      \def\colorgray#1{\color[gray]{#1}}%
      \expandafter\def\csname LTw\endcsname{\color{white}}%
      \expandafter\def\csname LTb\endcsname{\color{black}}%
      \expandafter\def\csname LTa\endcsname{\color{black}}%
      \expandafter\def\csname LT0\endcsname{\color{black}}%
      \expandafter\def\csname LT1\endcsname{\color{black}}%
      \expandafter\def\csname LT2\endcsname{\color{black}}%
      \expandafter\def\csname LT3\endcsname{\color{black}}%
      \expandafter\def\csname LT4\endcsname{\color{black}}%
      \expandafter\def\csname LT5\endcsname{\color{black}}%
      \expandafter\def\csname LT6\endcsname{\color{black}}%
      \expandafter\def\csname LT7\endcsname{\color{black}}%
      \expandafter\def\csname LT8\endcsname{\color{black}}%
    \fi
  \fi
    \setlength{\unitlength}{0.0500bp}%
    \ifx\gptboxheight\undefined%
      \newlength{\gptboxheight}%
      \newlength{\gptboxwidth}%
      \newsavebox{\gptboxtext}%
    \fi%
    \setlength{\fboxrule}{0.5pt}%
    \setlength{\fboxsep}{1pt}%
    \definecolor{tbcol}{rgb}{1,1,1}%
\begin{picture}(4648.00,3968.00)%
    \gplgaddtomacro\gplbacktext{%
      \csname LTb\endcsname
      \put(888,768){\makebox(0,0)[r]{\strut{}$0$}}%
      \csname LTb\endcsname
      \put(888,1122){\makebox(0,0)[r]{\strut{}$20$}}%
      \csname LTb\endcsname
      \put(888,1476){\makebox(0,0)[r]{\strut{}$40$}}%
      \csname LTb\endcsname
      \put(888,1830){\makebox(0,0)[r]{\strut{}$60$}}%
      \csname LTb\endcsname
      \put(888,2185){\makebox(0,0)[r]{\strut{}$80$}}%
      \csname LTb\endcsname
      \put(888,2539){\makebox(0,0)[r]{\strut{}$100$}}%
      \csname LTb\endcsname
      \put(888,2893){\makebox(0,0)[r]{\strut{}$120$}}%
      \csname LTb\endcsname
      \put(888,3247){\makebox(0,0)[r]{\strut{}$140$}}%
      \csname LTb\endcsname
      \put(1032,528){\makebox(0,0){\strut{}$1$}}%
      \csname LTb\endcsname
      \put(1828,528){\makebox(0,0){\strut{}$2$}}%
      \csname LTb\endcsname
      \put(2623,528){\makebox(0,0){\strut{}$4$}}%
      \csname LTb\endcsname
      \put(3419,528){\makebox(0,0){\strut{}$8$}}%
      \csname LTb\endcsname
      \put(4214,528){\makebox(0,0){\strut{}$16$}}%
    }%
    \gplgaddtomacro\gplfronttext{%
      \csname LTb\endcsname
      \put(3144,3064){\makebox(0,0)[r]{\strut{}Global}}%
      \csname LTb\endcsname
      \put(3144,2824){\makebox(0,0)[r]{\strut{}Neighbor-Only}}%
      \csname LTb\endcsname
      \put(228,2007){\rotatebox{-270.00}{\makebox(0,0){\strut{}Average Execution Time [s]}}}%
      \put(2623,168){\makebox(0,0){\strut{}Number of Nodes}}%
      \put(2623,3607){\makebox(0,0){\strut{}UTS}}%
    }%
    \gplbacktext
    \put(0,0){\includegraphics[width={232.40bp},height={198.40bp}]{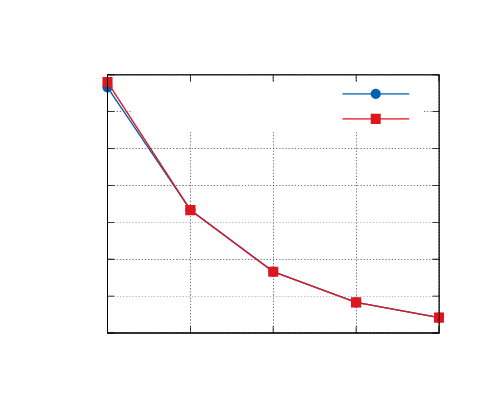}}%
    \gplfronttext
  \end{picture}%
\endgroup

%% file: plot_diff.tex
\begingroup
\bfseries
  \makeatletter
  \providecommand\color[2][]{%
    \GenericError{(gnuplot) \space\space\space\@spaces}{%
      Package color not loaded in conjunction with
      terminal option `colourtext'%
    }{See the gnuplot documentation for explanation.%
    }{Either use 'blacktext' in gnuplot or load the package
      color.sty in LaTeX.}%
    \renewcommand\color[2][]{}%
  }%
  \providecommand\includegraphics[2][]{%
    \GenericError{(gnuplot) \space\space\space\@spaces}{%
      Package graphicx or graphics not loaded%
    }{See the gnuplot documentation for explanation.%
    }{The gnuplot epslatex terminal needs graphicx.sty or graphics.sty.}%
    \renewcommand\includegraphics[2][]{}%
  }%
  \providecommand\rotatebox[2]{#2}%
  \@ifundefined{ifGPcolor}{%
    \newif\ifGPcolor
    \GPcolortrue
  }{}%
  \@ifundefined{ifGPblacktext}{%
    \newif\ifGPblacktext
    \GPblacktextfalse
  }{}%
  \let\gplgaddtomacro\g@addto@macro
  \gdef\gplbacktext{}%
  \gdef\gplfronttext{}%
  \makeatother
  \ifGPblacktext
    \def\colorrgb#1{}%
    \def\colorgray#1{}%
  \else
    \ifGPcolor
      \def\colorrgb#1{\color[rgb]{#1}}%
      \def\colorgray#1{\color[gray]{#1}}%
      \expandafter\def\csname LTw\endcsname{\color{white}}%
      \expandafter\def\csname LTb\endcsname{\color{black}}%
      \expandafter\def\csname LTa\endcsname{\color{black}}%
      \expandafter\def\csname LT0\endcsname{\color[rgb]{1,0,0}}%
      \expandafter\def\csname LT1\endcsname{\color[rgb]{0,1,0}}%
      \expandafter\def\csname LT2\endcsname{\color[rgb]{0,0,1}}%
      \expandafter\def\csname LT3\endcsname{\color[rgb]{1,0,1}}%
      \expandafter\def\csname LT4\endcsname{\color[rgb]{0,1,1}}%
      \expandafter\def\csname LT5\endcsname{\color[rgb]{1,1,0}}%
      \expandafter\def\csname LT6\endcsname{\color[rgb]{0,0,0}}%
      \expandafter\def\csname LT7\endcsname{\color[rgb]{1,0.3,0}}%
      \expandafter\def\csname LT8\endcsname{\color[rgb]{0.5,0.5,0.5}}%
    \else
      \def\colorrgb#1{\color{black}}%
      \def\colorgray#1{\color[gray]{#1}}%
      \expandafter\def\csname LTw\endcsname{\color{white}}%
      \expandafter\def\csname LTb\endcsname{\color{black}}%
      \expandafter\def\csname LTa\endcsname{\color{black}}%
      \expandafter\def\csname LT0\endcsname{\color{black}}%
      \expandafter\def\csname LT1\endcsname{\color{black}}%
      \expandafter\def\csname LT2\endcsname{\color{black}}%
      \expandafter\def\csname LT3\endcsname{\color{black}}%
      \expandafter\def\csname LT4\endcsname{\color{black}}%
      \expandafter\def\csname LT5\endcsname{\color{black}}%
      \expandafter\def\csname LT6\endcsname{\color{black}}%
      \expandafter\def\csname LT7\endcsname{\color{black}}%
      \expandafter\def\csname LT8\endcsname{\color{black}}%
    \fi
  \fi
    \setlength{\unitlength}{0.0500bp}%
    \ifx\gptboxheight\undefined%
      \newlength{\gptboxheight}%
      \newlength{\gptboxwidth}%
      \newsavebox{\gptboxtext}%
    \fi%
    \setlength{\fboxrule}{0.5pt}%
    \setlength{\fboxsep}{1pt}%
    \definecolor{tbcol}{rgb}{1,1,1}%
\begin{picture}(8502.00,5668.00)%
    \gplgaddtomacro\gplbacktext{%
      \csname LTb\endcsname
      \put(744,1186){\makebox(0,0)[r]{\strut{}$-4$}}%
      \csname LTb\endcsname
      \put(744,2022){\makebox(0,0)[r]{\strut{}$-2$}}%
      \csname LTb\endcsname
      \put(744,2858){\makebox(0,0)[r]{\strut{}$0$}}%
      \csname LTb\endcsname
      \put(744,3693){\makebox(0,0)[r]{\strut{}$2$}}%
      \csname LTb\endcsname
      \put(744,4529){\makebox(0,0)[r]{\strut{}$4$}}%
      \csname LTb\endcsname
      \put(888,528){\makebox(0,0){\strut{}$1$}}%
      \csname LTb\endcsname
      \put(2683,528){\makebox(0,0){\strut{}$2$}}%
      \csname LTb\endcsname
      \put(4478,528){\makebox(0,0){\strut{}$4$}}%
      \csname LTb\endcsname
      \put(6273,528){\makebox(0,0){\strut{}$8$}}%
      \csname LTb\endcsname
      \put(8067,528){\makebox(0,0){\strut{}$16$}}%
    }%
    \gplgaddtomacro\gplfronttext{%
      \csname LTb\endcsname
      \put(6998,4764){\makebox(0,0)[r]{\strut{}FIB}}%
      \csname LTb\endcsname
      \put(6998,4524){\makebox(0,0)[r]{\strut{}UTS}}%
      \csname LTb\endcsname
      \put(228,2857){\rotatebox{-270.00}{\makebox(0,0){\strut{}Performance Difference [\%]}}}%
      \put(4478,168){\makebox(0,0){\strut{}Number of Nodes}}%
    }%
    \gplbacktext
    \put(0,0){\includegraphics[width={425.10bp},height={283.40bp}]{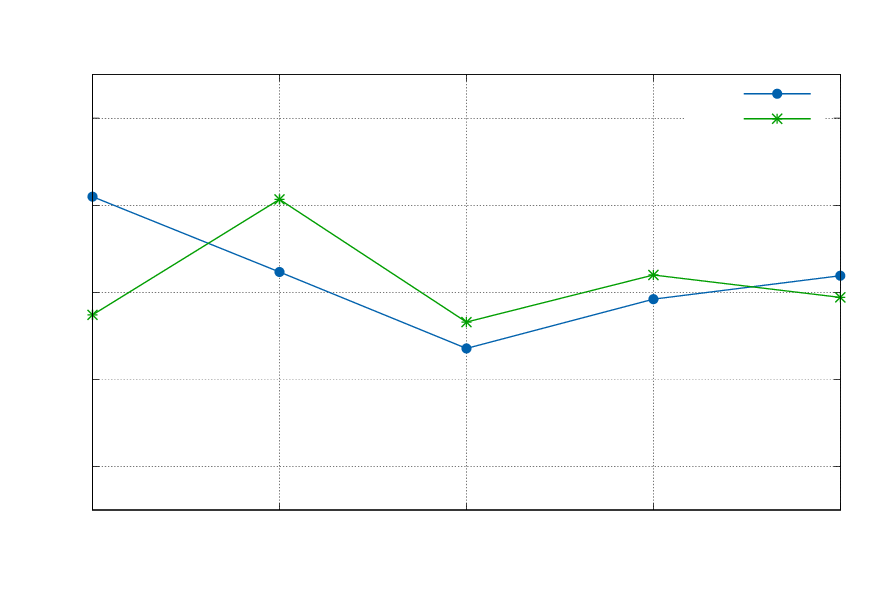}}%
    \gplfronttext
  \end{picture}%
\endgroup

%% file: 05relatedwork.tex
\section{Related Work}\label{sec:relatedwork}

Satellite missions have traditionally run pre-programmed command sequences uploaded from ground stations.
Recent SEC research proposes load balancing by forwarding tasks through the network~\cite{SECloadbalancing} or migrating entire program runs to satellites with available resources~\cite{OrbitalEdgeComputing}.
These approaches schedule whole program runs, whereas AMT work stealing operates at task granularity and reacts to load imbalances during execution.

Blumofe and Leiserson proved that random work stealing is optimal for fully connected networks with uniform latency~\cite{CilkWS}.
Their analysis assumes that every steal attempt has equal cost regardless of which worker is targeted.
This assumption does not hold for topologies with non-uniform communication costs, such as LEO constellation meshes, where multi-hop steals are much more expensive than single-hop ones.

Several topology-aware strategies address this non-uniform latency.
Hierarchical work stealing prioritizes nearby workers before attempting remote steals~\cite{HotSLAW}.
In the lifeline scheme, a thief tries a fixed set of preferred targets before attempting random global steals~\cite{Lifeline}.
Both approaches retain global stealing as a fallback when local attempts fail.
In contrast, our proposed neighbor-only strategy does not fall back to global steals, keeping all steal communication within a single hop.
We make this choice because ISL per-hop latency is high: even two-hop steals are much more expensive than single-hop ones.

Locality-aware scheduling has also been studied in the context of Non-Uniform Memory Access~(NUMA) architectures.
On multi-socket and multi-node HPC systems, NUMA-aware work stealing first selects victims that share the same local memory to reduce remote memory access overhead~\cite{HotSLAW}.
Restricting steals to nearby workers can improve performance even though it reduces the set of potential victims, but NUMA hierarchies have only two or three levels, while distances in a satellite mesh grow with the square root of the constellation size.
HPC clusters also use structured network topologies such as dragonfly or fat-tree, but latency differences across hops are much smaller than in a satellite mesh.

Terrestrial edge computing also distributes computation across geographically dispersed nodes~\cite{Shi2016EdgeVisionChallenges,Mao2017MECSurvey}, but SEC faces different constraints.
Terrestrial edge nodes have reliable power supplies, persistent storage, and low-latency links to cloud data centers.
SEC satellites, by contrast, depend on solar power with periodic shadow-induced outages, have limited onboard storage, and communicate through ISLs with latencies that scale with hop count.
Scheduling strategies from terrestrial edge computing therefore do not directly apply to SEC.

In HPC, process failures are typically handled through Checkpoint/Restart~(C/R), which periodically saves the full application state to persistent storage~\cite{SurveyCRdistr}.
SEC satellites can also fail due to hardware faults, but their limited storage, lack of a shared file system, and high-latency links make traditional C/R impractical.
Task-Level Checkpointing~(TC) is an alternative that saves only pending tasks and the intermediate results needed to continue execution, reducing the checkpoint size compared to full C/R.
Most TC approaches replicate checkpoints across multiple nodes for redundancy~\cite{PosnerFT20,JonasIJNC22}, though some avoid replication entirely~\cite{MarcoFTJSEA}.
As a third option, supervision lets the victim track which tasks were stolen from it and re-execute them if the thief fails~\cite{KrishnaNestedFJournal}.
In HPC, supervision minimizes overhead during failure-free execution, while TC offers faster recovery at scale~\cite{JonasIJNC22}.
In SEC, communication latency is higher and satellites face failure modes that do not exist in HPC, such as radiation-induced failures and power loss in the shadow of Earth.

Some shutdowns are predictable, such as battery depletion before a satellite enters the shadow of Earth.
Malleable AMT programs handle predictable shutdowns by adding or removing workers at runtime without interrupting the computation~\cite{JonasElasticityWS,PosnerDPP25}.
A malleable runtime could therefore let a satellite migrate its tasks and results to neighbors before powering down.
Our experiments used a static 2D grid topology, but combining neighbor-only stealing with malleability is a natural extension for adapting to predictable topology changes during execution.

%% file: 06conclusions.tex
\section{Conclusions}\label{sec:conclusions}
This paper proposed a neighbor-only work stealing strategy for AMT runtimes on LEO satellite mesh topologies and evaluated it through an analytical model and preliminary experiments.
Our contribution has two parts.
First, preliminary experiments with the ItoyoriFBC runtime on up to~640 cores of an HPC cluster with a uniform low-latency interconnect and an emulated 2D mesh show that our proposed neighbor-only strategy performs within~$\pm 2.2\,\%$ of global stealing for both the balanced FIB and the irregular UTS workloads.
This suggests that, in this low-latency control setting, restricting the victim set to neighbors does not noticeably reduce how often steal attempts succeed.
Second, our analytical model suggests that, once realistic ISL hop latency is taken into account, neighbor-only stealing gains a per-attempt latency advantage over global stealing that grows with the square root of the constellation size.
Taken together, the preliminary experiments support the assumption used by the model, while the model suggests that neighbor-only stealing becomes preferable at scale.

Our experiments deliberately isolated victim selection from ISL latency; empirical evaluation on an emulated high-latency mesh and on real SEC applications is future work.
Further directions include forwarding result messages along the mesh topology and gradually considering more distant victims after consecutive failed steal attempts.